\documentclass[12pt,epsf]{article}

\usepackage{amsmath,amssymb,graphicx}

\newcommand{\beq}{\begin{eqnarray}}% can be used as {equation} or  {eqnarray}
\newcommand{\eeq}{\end{eqnarray}}

\newcommand{\be}{\begin{equation}}
\newcommand{\ee}{\end{equation}}
\newcommand{\bear}{\begin{eqnarray}}
\newcommand{\eear}{\end{eqnarray}}

\newcommand{\cO}{{\cal O}}

\newcommand{\Frac}[2]{\frac{\displaystyle #1}{\displaystyle #2}}
\newcommand{\Int}{\displaystyle{\int}}
\title{
{\huge \bf Higgs decay into diphoton in the Composite Higgs Model } \vspace*{0.8cm}
 }

\author{Haiying Cai   \\
 \\ \normalsize\emph{Department of Physics, Peking University, Beijing 100871, China}  \\ }

\date{}   %{\today}
\begin{document}
\setcounter{page}{0} \maketitle
\thispagestyle{empty}
\vspace*{0.5cm} \maketitle  %\begin{center}

\begin{abstract}
We explore the Higgs couplings to  gauge bosons in the minimal
$SO(5)/SO(4)$ 4D composite Higgs model. The pion scatterings put
unitary constraints on the couplings and therefore determine the
branching ratios of  various Higgs decays. Through fine-tuning the
parameters, enhancement of  Higgs to diphoton rate is possible to
be achieved with the existence of  vector meson fields.
\end{abstract}

\thispagestyle{empty}

\newpage

\setcounter{page}{1}

\section{Introduction}
The composite Higgs model provides an alternative solution to the little hierarchy problem compared with the well-known supersymmetric models, since the economic formulation of the standard model is impossible to explain the lightness of  Higgs mass. The composite Higgs boson emerges as a pseudo-Nambu-Goldstone boson (pNGB)  from a spontaneously broken  global symmetry,  therefore its mass is much lighter than the other resonances from the strong dynamic sector. The original minimal composite model  is realized in the five-dimensional  Randall Sundrum model, and  the Higgs is the fifth component of  the broken gauge bosons~\cite{holographic}. Using the holographic approaching,  the effective Lagrangian is gained after integrating out the bulk field  with the UV brane value fixed.  The potential for the holographic composite Higgs could be calculated in the form of brane to brane 5D propagators~\cite{propagator}. In the last few years increased attention has been focused on the deconstruction version of the 5D theory, which leads to varieties of  4D composite Higgs models, assuming the existence of one elementary sector and one strong interaction sector~\cite{CH}.  Without the presence of  additional composite fields,  the  composite Higgs has a reduced coupling with the gauge bosons, which may lead to the violation of unitary  in the pion scatterings before the cutoff scale is reached.  The method to restore the perturbative unitary  is to introduce vector resonances. The unitary requirement will correlate the global symmetry breaking scale $f$ with the mass $m_\rho$ of the vector resonance. It is interesting that the presence of the vector resonance will also modify the Higgs coupling, with its deviation parametrized by $\xi = v^2/f^2$, which in turn changes the branching ratio of various Higgs decay.  Another crucial ingredient in the composite Higgs model is the partial compositeness of gauge bosons due to the nonlinearity, with the degree of compositeness mainly controlled by the gauge couplings. In this paper we  first review a simple  model setup of  4D composite Higgs and show that  it is capable to accommodate the $125 ~\mbox{GeV} $ resonance with the appropriate properties recently discovered at the LHC~\cite{Higgs}.

\section{Lagrangian of the sigma model}
Let us start with the basic model setup.  Our Higgs is realized
as one pNGB from a strong interacting sector using the nonlinear
sigma model. We formulate the effective Lagrangian for those pNGBs
via the Callan-Coleman-Wess-Zumino (CCWZ) prescription~\cite{CCWZ}. 
In the following, we are going to review the nonlinear realization of composite Higgs and
capture the necessary ingredients for our calculations.
Considering the global symmetry breaking pattern $ SO(5) \to
SO(4)$,  there are four  pNGBs which fit a basic representation
of the $SO(4)$ symmetry group.  The first three, i.e.
$\pi_{1,2,3}$,  are eaten by the $W,Z $ bosons, with the remaining
one, $\pi_4$, identified as the Higgs.  Denoting the Goldstone
bosons as $U = exp (i \sqrt 2 \pi^{\hat a} T^{\hat a}/f)$,  the
sigma field would transform nonlinearly under the  full global
symmetry as $ U \to  g U h^\dagger (g, \pi) $, and one can
calculate the structure of $ i U^\dagger \partial_\mu U =
d_\mu^{\hat a} T^{\hat a}  + E_\mu^{a_L} T^{a_L} + E_\mu^{a_R}
T^{a_R} $, where $T^{\hat a}, \hat a = 1,2 ,3 ,4 $ are the broken
generators in the coset space of  $SO(5)/SO(4)$, and $T^{a_L(a_R)},
a_L(a_R) = 1,2 ,3$ are the unbroken generators in the $SO(4)\simeq
SU(2)_L \times SU(2)_R $ symmetry group. It should be noted that
all the generators in the $SO(5)$ symmetry group are normalized as
$\mbox{Tr} (T^a T^b) = \delta^{a b}$. The building blocks of the
CCWZ formalism are the variables $d_\mu^{\hat a}$ and $E_\mu^{a_L,
a_R}$ decomposed in the broken and unbroken generator directions
respectively.  Following the usual formulation, we  gauge a subgroup $SU(2) \times U(1) $ in
the global $SO(4)$, resulting in an explicit breaking of the full
global symmetry. The gauged CCWZ structures are calculated in a
similar approach by substituting $\partial_\mu$ with the covariant
operator $D_\mu = \partial_\mu -i g_0 W_\mu^a T_L^a - i B_\mu
T_R^3 $.  At the leading order of the chiral expansion,
$d_\mu^{\hat a}$ and $E_\mu^{a_L, a_R}$ are expressed as:
\beq
d_\mu ^{\hat a} &=&  - \frac{\sqrt 2 }{f} D _\mu \pi ^{\hat a} + \frac{ \sqrt 2}{3 f^3}{\left[ \pi ,\left[ \pi, D_\mu \pi  \right]\right]^{\hat a}} + \cdots  \,, \\
E_\mu ^a &=&  g_0 W_\mu ^a + {g'}_0 B_\mu \delta ^{a3} + \frac{i}{f^2}{\left[ \pi , D_\mu \pi \right]^a} +\cdots
\eeq
under the local symmetry group,  the corresponding transformation rules are:
\beq
d_\mu \to  h(g, \pi ) d_\mu h^\dagger (g, \pi) ,  \quad  E_\mu \to h(g, \pi) E_\mu h^\dagger (g,\pi) + i h(g, \pi) \partial_\mu h^\dagger  (g, \pi)
\eeq
since $E_\mu$ behaves as a gauge field, the coupling of Goldstone
bosons to the fundamental fermions is via the covariant derivative
$\partial_\mu - i E_\mu$. In this paper we are only concerned with
the vector meson effects and  would not explore too much into the
fermion sector. We can conveniently calculate the mass terms for
the $W$ and $Z$ gauge bosons after   electroweak symmetry breaking
through the kinetic terms:
\beq
 \frac{f^2}{4} \mbox{Tr} ~ {d^\mu}{d_\mu} = \frac{1}{2} \frac{2m_W^2}{v} ( v+ 2 a h + b \frac{h^2}{v}) W_\mu^+ W_\mu^-  + \frac{1}{2} \frac{m_Z^2}{v} (v+2 a h + b \frac{h^2}{v}) Z_\mu Z_\mu  + \mathcal{O}(h^3)~
\eeq
\beq
 m_W^2 = \frac{g_0^2 {f^2}\sin {\theta ^2}}{4}, ~ m_Z^2 = \frac{( g_0^2 + g_0^{\prime 2} ){f^2}\sin {\theta ^2}}{4},  ~ a = \cos \theta , ~ b= \cos^2 \theta -\sin^2 \theta\eeq
where the parameters $a$ and $b$,  both of which are always less
than one, indicate that the  Higgs couplings are reduced as
compared with the Standard Model. In the minimal $SO(5)/SO(4)$ setup,  $\theta $ is the misalignment of the true vacuum relative to
the gauged $SO(4)$ subgroup, with the VEV of the Higgs  defined as
$v = f \sin \theta = 246.0 ~\mbox{GeV} $.

Under the  partial UV completion hypothesis~\cite{Contino:2011np},  one pair of $\rho_{L \mu } $ and $\rho_{R \mu} $ in the representations  $(3,1) \oplus (1,3)$ of $SU(2)_L \times SU(2)_R$,  transforming under the local symmetry group  as $\rho \to h(g, \pi) \rho_\mu h^\dagger(g, \pi) + i h(g, \pi) \partial_\mu h ^\dagger(g, \pi) $, needs to be added into the strong dynamic sector. The gauge invariant Lagrangian for the vector resonances consisting of  kinetic terms and mass terms is formulated as:
\beq
\mathcal{L}_{\rho L} &=&  - \frac{1}{4} \mbox{Tr} \left( {{\rho _{L,\mu \nu }}\rho _L^{\mu \nu }} \right) + \frac{{a_{\rho_{L}}^2 f^2}}{2}  \mbox{Tr} {\left( {{g_{\rho_{L}}}{\rho _{L\mu }} - E_\mu ^L} \right)^2}    \label{rhoL}  \,, \\
\mathcal{L}_{\rho R}  &=&  - \frac{1}{4} \mbox{Tr} \left( {{\rho _{R,\mu \nu }}\rho _R^{\mu \nu }} \right) + \frac{{a_{\rho_{R}}^2 f^2 }}{2} \mbox{Tr}  {\left( {{g_{\rho_{R}}}{\rho _{R\mu }} - E_\mu ^R} \right)^2}   \,. \label{rhoR}
\eeq
At the low energy scale, we are only interested in the interactions which are relevant to the pion scatterings, that is  the  Goldstone bosons self-interactions and at most their interactions with the vector resonances. After a little bit of algebra,  it is easy to reach the explicit  Lagrangian:
\beq
\mathcal{L}_{\rho_L \pi^2+\pi^4}& = &  \frac{{a_{\rho L}^2{g_{\rho L}}}}{2}\left[ {{\varepsilon ^{ijk}}{\pi ^i}{\partial _\mu }{\pi ^j}\rho _{L\mu }^k + \left( {{\pi ^k}{\partial _\mu }{\pi ^4} - {\pi ^4}{\partial _\mu }{\pi ^k}} \right)\rho _{L\mu }^k} \right]    \nonumber  \\
&&  - \frac{{a_{\rho L}^2}}{{8{f^2}}}\left[ {{{\left( {{\pi ^a}{\partial _\mu }{\pi ^a}} \right)}^2} - {{\left( {{\pi ^a}{\partial _\mu }{\pi ^b}} \right)}^2}} \right]\\
\mathcal{L}_{\rho_R \pi^2+\pi^4}  &=&   \frac{{a_{\rho R}^2{g_{\rho R}}}}{2}\left[ {{\varepsilon ^{ijk}}{\pi ^i}{\partial _\mu }{\pi ^j}\rho _{R\mu }^k - \left( {{\pi ^k}{\partial _\mu }{\pi ^4} - {\pi ^4}{\partial _\mu }{\pi ^k}} \right)\rho _{R\mu }^k} \right]   \nonumber \\
&&  - \frac{{a_{\rho R}^2}}{{8{f^2}}}\left[ {{{\left( {{\pi ^a}{\partial _\mu }{\pi ^a}} \right)}^2} - {{\left( {{\pi ^a}{\partial _\mu }{\pi ^b}} \right)}^2}} \right]
\eeq
Since the $h \pi^2 $ and $h^2 \pi^2$  interactions are  determined by $a$  and $b$, whereas the pion self-interaction and pion interaction with vector meson are related to $ a_\rho$ and the global symmetry breaking scale $f$, the correlation between those parameters and the allowed parameter space information could be extracted from both the pion elastic and the pion inelastic scatterings.

\section{ Consider Elastic and Inelastic Pion  Scattering}

\subsection{$\pi\pi\to\pi\pi$ scattering}

\begin{figure}[!t]
\begin{center}
\includegraphics[angle=0,clip,width=7.5cm]{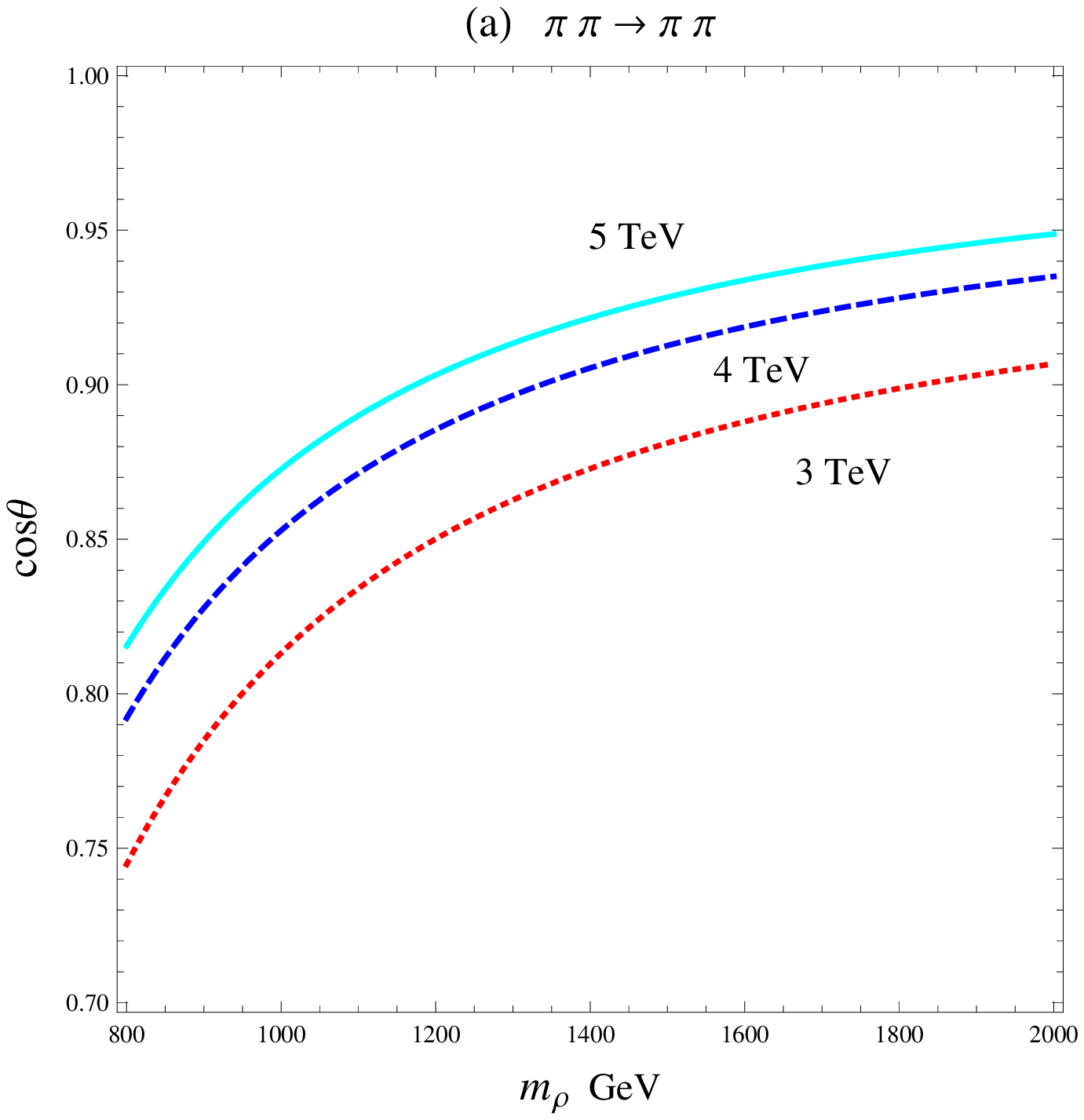}
\includegraphics[angle=0,clip,width=7.5cm]{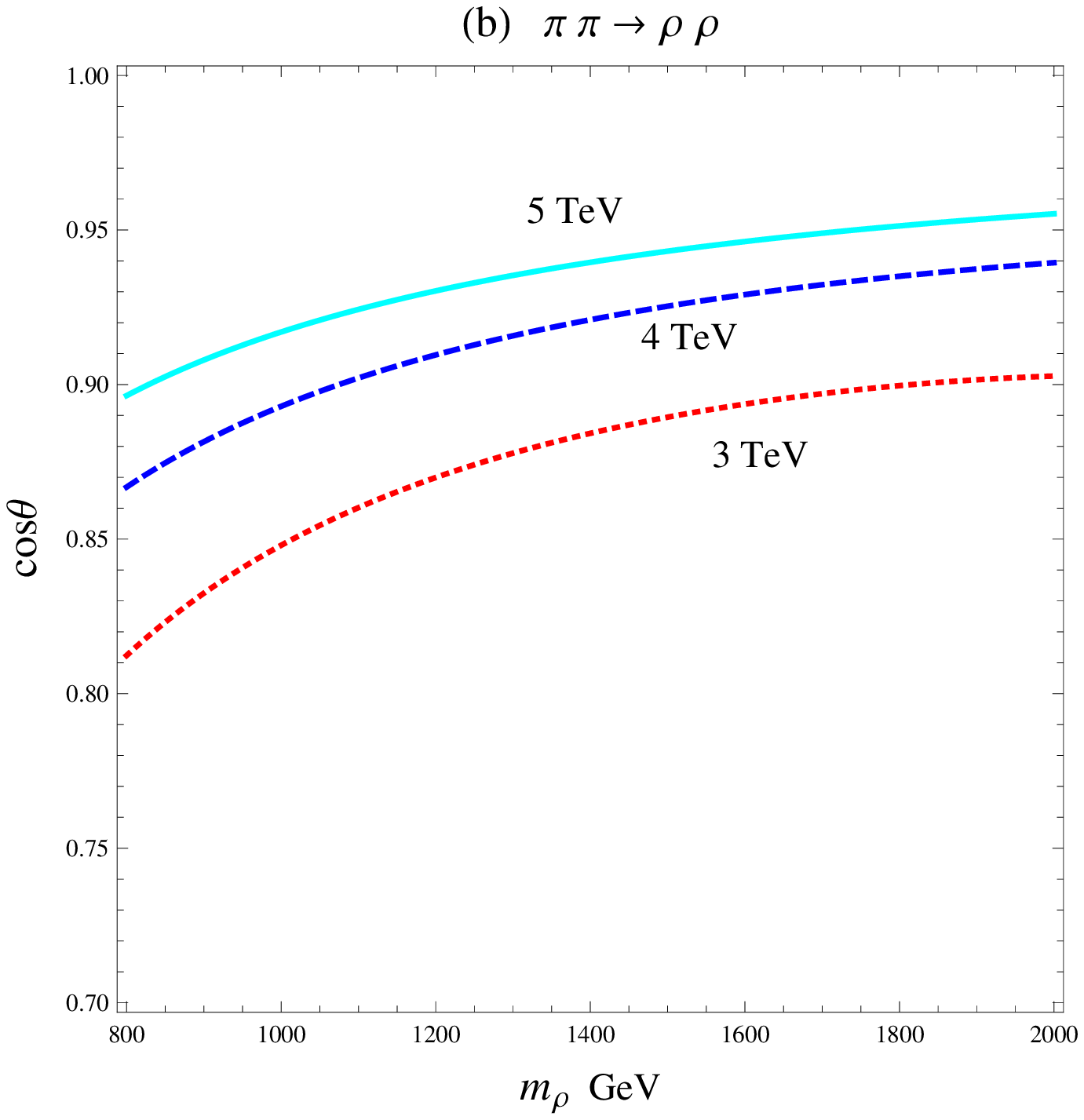}
\caption{{\small Parameter-space region where the unitarity bound
$|$Re$a^0_0(s)^{(\pi\pi)}|\leq \frac{1}{2}$ is violated at
energies $s\leq \Lambda^2$, for $\Lambda=3.0$~TeV (red dotted
line), $\Lambda=4.0$~TeV (blue dashed line) and $\Lambda=5.0$~TeV
(cyan solid line), with $g_\rho =2.0$, i.e.  only the regions above the lines are permitted by perturbative unitary. The left panel is for the
pion elastic scattering and the right panel  is for the pion
inelastic scattering. }} \label{fig.pipi-bound-region}
\end{center}
\end{figure}

For the scattering $\pi^a \pi^b \to \pi^c  \pi^d$ of the $SU(2)$-triplet Goldstones, the amplitude has the general isospin structure:
\beq
 A({\pi ^a}{\pi ^b} \to {\pi ^c}{\pi ^d}) = A(s,t,u)^{(\pi \pi )} \delta ^{ab} \delta ^{cd} + A(t,s,u)^{(\pi \pi )} \delta ^{ac} \delta ^{bd} + A(u,t,s)^{(\pi \pi )} \delta ^{ad} \delta ^{bc}
\eeq
\beq
&& A(s,t,u)^{(\pi \pi)} = \frac{s}{v^2} - \frac{a_{\rho L} ^2}{4 f^2}\left[ {3s + m_{\rho L} ^2\left( {\frac{{s - u}}{{t - m_{\rho L} ^2}} + \frac{{s - t}}{{u - m_{\rho L} ^2}}} \right)} \right]  \nonumber \\
&& \qquad - \frac{a_{\rho R} ^2}{4 f^2}\left[ {3s + m_{\rho R} ^2\left( {\frac{s - u}{t - m_{\rho R} ^2} + \frac{s - t}{u - m_{\rho R} ^2}} \right)} \right] - \frac{a^2}{v^2}\frac{s^2}{s - m_h^2}
\eeq
where the terms with dependence on the mass $m_{\rho_{L,R}}$ comes
from $\rho_{L,R}$ meson mediated $t$ channel and $u$ channel
diagrams,  and the last term comes from the light Higgs mediated
$s$ channel diagram, whereas the remaining terms come from the
contact interaction. The amplitude can be decomposed into $1,3, 5
$ in the isospin basis: \beq
T_0(s,t,u) &=&  3 A(s, t, u ) + A(t, s, u) + A(u,t, s) , \\
T_1(s,t,u) &=& A(t, s, u) - A(u, t, s), \\
T_2(s,t,u) &=& A(t, s, u)+ A(u, t, s)\, .
\eeq
It is then possible to transform these isospin amplitudes in terms of the partial wave (PW) decomposition
\bear
T^I(s,t) \,=\, \sum_{J=0}^\infty 32\pi \, (2J+1) \, P_J(\cos\alpha) \, a^I_J(s)\, ,
\eear
with the partial waves provided by
\bear
a^I_J(s) \,=\, \Frac{1}{64\pi} \, \Int_{-1}^{+1} {\rm d}\cos\alpha\,
P_J(\cos\alpha) \, T^I(s,t(s,\cos\alpha))\, ,
\label{eq.PW-projection}
\eear
and with $\cos\alpha=1- 2t/s$.  In this normalization the partial waves  can be written in the
form $a^I_J(s)= \frac{i}{2}(1-\eta e^{2i\delta^I_J})$, with the inelasticity obeying
the unitarity bound $0\leq \eta\leq 1$. This implies the constraints
\bear
|\mbox{Re}a^I_J(s)|\, \leq\, \frac{1}{2}\, , \qquad\qquad
\mbox{Im}a^I_J(s)\, \leq 1\, , \qquad\qquad
|a^I_J(s)|\, \leq \, 1\, .
\label{eq.PW-unitarity}
\eear
We will make use of the first one in order to constrain  our partial wave amplitudes.
One must be aware of the slight arbitrariness of this choice,  as we
could also consider the last constraint in~(\ref{eq.PW-unitarity})
to determine when the theoretical determinations ``violate'' unitarity.
The root of this ambiguity lies on the fact that the tree-level amplitude
is never truly unitary for $s>0$, as the tree-level PW always lies out of the
Argand circle and has inelasticity $\eta>1$.
Nevertheless,  as far as $|$Re$a^I_J(s)|\leq 1/2$  it is still possible to argue
that our perturbative tree-level estimate 
still provides a good approximation of  the full amplitude.
In the light Higgs limit $m_h^2\ll | s |$ we get the partial wave power expansion,
\bear
a^0_{0}(s)^{(\pi\pi)}
 &=&
\Frac{(4 - 3 a_{\rho L}^2 - 3 a_{\rho R}^2 ) \, s}{64 \pi f^2}
\nonumber\\
&&
\, + \,
\bigg[    \, \Frac{a_{\rho L}^2 m_{\rho L}^2
\left(\left(\Frac{m_{\rho L}^2}{s}+2 \right)
   \log \left(\Frac{s}{m_{\rho L}^2}+1\right)-1\right)}{32 \pi   f^2}
   \,+\, ( L \leftrightarrow R) \, \bigg]\, .
\eear
Notice that  we have made use of the $SO(5)/SO(4)$ relations $v=f
\sin\theta$ and $a=\cos\theta$. The first term in the r.h.s. of
the equation diverges like $\sim \cO(s)$ at high energies and
spoils that unitarity bound very quickly. Hence, one usually
requires the exact cancellation of the $\cO(s)$  term in the
high-energy $\pi\pi$ scattering~\cite{Marzocca,Csaki,Guo}, this
is,
\bear
a_{\rho L}^2 \, +\, a_{\rho R}^2   &=& \Frac{4}{3}\, .
\eear
For the left-right symmetric case $ a_{\rho L} = a_{\rho R} =
a_\rho $ this turns into \bear a_\rho^2  &=& \Frac{2}{3}\, . \eear 
It should be noticed that after imposing $a_\rho^2=2/3$  the
partial wave amplitude behaves like $a^0_0(s)\simeq
\frac{m_\rho^2}{24 \pi f^2} \bigg(2\ln\frac{s}{m_\rho^2}-1\bigg)$
at high energies. However, this mild $\ln{(s)}$  divergent
behavior at high energies will eventually exceed the
``unitarity'' bound.   Since the mass of vector resonance is
$m_\rho = a_\rho  g_\rho f$ , as we fix the coupling $g_\rho$, two independent
parameters are left. We are going to adopt another method to
constrain the parameter space of $(a, m_\rho)$ by demanding the
unitary bound is satisfied below a fixed cutoff scale $\Lambda$.
In Fig.~\ref{fig.pipi-bound-region} (a), we have plotted the
parameter-space region where the unitarity bound
$|$Re$a^0_0(s)^{(\pi\pi)}|\leq  \frac{1}{2}$ is violated at
energies $s\leq \Lambda^2$, for $\Lambda=3.0$~TeV (red dotted
line),
 4.0~TeV (blue dashed line), and 5.0~TeV (cyan solid line).
It is interesting to observe that as  the resonance mass
grows, the allowed region where perturbation theory is applicable
(or, conversely, where the ``unitarity--bound'' is satisfied)
gets more and more reduced.
\subsection{$\pi\pi\to h h$ scattering}

For the inelastic scattering: $A({\pi ^a}{\pi ^b} \to {h}{h}) = A{(s,t,u)^{(hh)}}{\delta ^{ab}}$. The isospin structure is quite simple in this process and  it gets a contribution from the contact interaction, along with $\pi$, $ \rho_L$, and  $\rho_R$ exchanged  $t $ and $ u$ channels. The full prediction is:
\beq
&& A(s,t,u)^{(hh)} = -\frac{b}{v^2}s - \frac{a^2}{v^2}\left[ {\frac{\left( {t - m_h^2} \right)^2}{t} + \frac{\left( {u - m_h ^2} \right)^2}{u}} \right]  + \frac{a_{\rho L}^2 +a_{\rho R} ^2  }{4 f^2}(- 3s + 2 m_h^2)  \nonumber \\ && \qquad - ~\frac{a_{\rho L} ^2 }{4 f^2 } m_{\rho L} ^2 \left( \frac{s - u}{t - m_{\rho L} ^2} + \frac{s - t}{u - m_{\rho L} ^2} \right)
- \frac{a_{\rho L} ^2 }{4 f^2 }  m_h^4 \left( \frac{1}{t - m_{\rho L} ^2} + \frac{1}{u - m_{\rho L} ^2} \right)  \nonumber \\
&& \qquad  - ~\frac{a_{\rho R} ^2 }{4 f^2 } m_{\rho R} ^2 \left( \frac{s - u}{t - m_{\rho R} ^2} + \frac{s - t}{u - m_{\rho R} ^2} \right)
- \frac{a_{\rho R} ^2 }{4 f^2 }  m_h^4 \left( \frac{1}{t - m_{\rho R} ^2} + \frac{1}{u - m_{\rho R} ^2} \right)
\eeq
One may then perform a PW projection similar to that in
Eq.~(\ref{eq.PW-projection}) but with the effect of Higgs mass included in  $\cos\alpha=2 (t -m_h^2+s/2)/(s\beta_h(s))$ and the phase-space factor $\beta_h(s)=\sqrt{1- 4 m_h^2/s}$.
In the light Higgs limit $m_h^2\ll |s|$ one gets
\bear
a^0_0(s)^{(hh)} &=&  \Frac{1}{2}\, a^0_0(s)^{(\pi\pi)} \, .
\eear
Here we made use of $a=\cos\theta$ and $b=\cos 2\theta$.   With
the existence of  $SO(5)$ global symmetry, we expect to get the
same expectation as in $\pi\pi\to\pi\pi$  when we demand the
$\cO(s)$ term to exactly cancel out at high energies. Nonetheless,
due to the extra factor $\frac{1}{2}$,  the violation of our
``unitarity bound'' by the linear $s$ divergence and  the residual
$\ln(s)$ high energy divergence occurs later.

\subsection{$\pi\pi\to\rho\rho$ scattering}

Now we come to consider the inelastic scattering $\pi^a \pi^b \to
\rho_L^c \rho_L^d $, where  the longitudinal component of the
vector meson is parametrized as $\epsilon_L (k)= \bigg(
\frac{\sqrt s}{2} \frac{\beta_\rho}{m_\rho},  \frac{\sqrt s}{2
m_\rho} \vec{n}_k\bigg) $ with $\beta_\rho = \sqrt{1-4
m_\rho^2/s}$. As the Higgs is realized as one pNGB,  the  Higgs
coupling $ \frac{c_\rho}{f^2} m_\rho^2 v h \rho_\mu^a \rho_\mu^a $
comes from the mass term of the vector meson  and the parameter
$c_\rho$ is suppressed by $g_0^2/g_\rho^2$. The isospin
decomposition is similar to the elastic one but with two form
factors. For  $A(s,t, u)$  three diagrams contribute: the $s$
channel $h^0$ exchanged diagram,  $t$ channel and $u$ channel
$\pi$ exchanged diagrams;  whereas for $B(s, t, u)$, there is one
$\rho_\mu$  mediated  $s$ channel diagram,  one $\pi$ mediated $u$
channel diagram, and one $h^0$ mediated  $t$ channel  diagram:
\beq
A({\pi ^a}{\pi ^b} \to {\rho_L ^c}{\rho_L ^d}) = A(s,t,u)^{(\rho \rho)}{\delta ^{ab}}{\delta ^{cd}} + B(s,t,u)^{(\rho \rho)}{\delta ^{ac}}{\delta ^{bd}} + B(s,u,t)^{(\rho \rho)}{\delta ^{ad}}{\delta ^{bc}},
\eeq
\beq
A(s,t,u) &=& \frac{ a c_\rho}{f^2}\frac{s (s- 2 m_\rho^2)}{s-m_h^2}  + \frac{a_{\rho} ^2 }{4 f^2 \beta_\rho^2 } \frac{1}{u } \left(  \frac{s}{2}\left(\beta_\rho^2 +1\right)+t -m_\rho^2\right)^2 \nonumber \\
& + & \frac{a_{\rho} ^2 }{4 f^2\beta_\rho^2} \frac{1}{t} \left(  \frac{s}{2}\left(\beta_\rho^2 -1\right) -t + m_\rho^2\right)^2 ~,    \\
B(s,t,u) &=& \frac{1}{4 f^2 }\frac{(s+2 m_\rho^2)(t - u)}{(s - m_{\rho} ^2)} + \frac{a_{\rho} ^2 }{4 f^2 \beta_\rho^2 } \frac{1}{u } \left(  \frac{s}{2}\left(\beta_\rho^2 +1\right)+t -m_\rho^2\right)^2 \nonumber \\
& + & \frac{a_{\rho} ^2 }{4 f^2\beta_\rho^2} \frac{1}{(t-m_h^2)} \left(  \frac{s}{2}\left(\beta_\rho^2 -1\right) -t + m_\rho^2\right)^2 ~.
\eeq
Since the vector resonance is introduced to restore the
perturbative unitary,  it is usually demanded that the cutoff
scale satisfying $ 2 m_\rho  < \Lambda <  4 \pi f $. When the threshold effect of the final states could be ignored, i.e. $m_\rho \ll \Lambda$,   there are only linear growing $s$ term and constant term  in the partial wave transformation. But as the mass $m_\rho $ is comparable with $\Lambda$,  logarithmic terms also appear:
\bear
&& a^0_0(s)^{(\rho \rho)} =  \Frac{\left(24 a  c_\rho \beta_\rho^3+10 a_\rho^2 \left(1-2 \beta_\rho^2\right) \beta_\rho +5 a_\rho^2 \left(1- \beta_\rho^2\right)^2\log \left(\frac{(1-\beta_\rho ) s-2  m_\rho^2}{(1+ \beta_\rho ) s-2 m_\rho^2}\right) \right) s }{256 \pi f^2 \beta_\rho^3 } \nonumber \\ 
&& \quad  - \Frac{\left(  \left(12 a c_\rho \beta _\rho^2 +5 a_\rho^2\right) \beta_\rho + 5 a_\rho^2 \left(\left(1- \beta_\rho ^2\right)  -\frac{m_\rho^2 }{s} \right)  \log \left(\frac{(1-\beta_\rho ) s-2  m_\rho^2}{(1+ \beta_\rho ) s-2 m_\rho^2}\right)\right) m_\rho^2}{64 \pi f^2 \beta_\rho^3 }
\eear
In the limit $m_h, m_\rho \ll \Lambda $,  the partial wave displays a linear growing pattern: $a^0_0(s)^{(\rho \rho)} \simeq \Frac{s}{128 \pi f^2}{\bigg(12 a c_\rho - 5 a_\rho^2\bigg)}$, which pushes the partial wave to grow quickly after the two mesons threshold is reached, and this process  provides a
complementary constraint for the parameter space. In
Fig.~\ref{fig.pipi-bound-region} (b), with the $c_\rho$ term
ignored, we show that the unitary bound  actually imposes a more stringent
constraint on the parameter space, that is it requires one larger
value of  global symmetry breaking scale $f$ (i.e. $\cos \theta$
needs to be more close to 1) for the same value of $m_\rho$ and
$\Lambda$ as compared with the elastic scattering.

\section{Higgs to diphotons from $\rho$ mesons}

In this section, we discuss the resonance effects on the Higgs
sector.  The gauge bosons couple to the vector resonances via the
mass terms described in Eq.(\ref{rhoL}-\ref{rhoR}),  since it is 
the combination of $(g_\rho \tilde{\rho}_\mu^{L,R} - E_\mu^{L,R})$
that transforms homogeneously under the symmetry group of $SO(4)$. Due
to the mixing between the gauge eigenstates of  $\tilde{\rho}_{L
\mu}^a$, $\tilde{W}_\mu^a$, $\tilde{\rho}_{R \mu}^3$ and $
\tilde{B}_\mu$, the exact gauge bosons in the effective theory
gain the property of partial compositeness. At the leading order
of $\xi = v^2 /f^2= (1-\cos^2 \theta) $,  the mass terms are
simplified as:
\beq
\mathcal{L}_{\rho }^{m}
&=&  \frac{m_\rho^2}{2 g_\rho^2} \left( g_{\rho}\tilde{\rho} _{L\mu }^a - g_0 \tilde{W}_\mu^a \right)^2  + \frac{m_\rho^2}{2 g_\rho^2} \left( g_{\rho} \tilde{\rho} _{R\mu }^3 - g_0^\prime \tilde{B}_\mu \right)^2
\eeq
such that the gauge couplings for  $SU(2)_L \times U(1)_Y$ in the standard model are determined by the relations of:
\beq
 g_2^2 = \frac{g_0^2g_{\rho} ^2}{{g_0^2 + g_{\rho} ^2}} ~,  \quad  g_1^2 = \frac{{g_0^{\prime 2}g_{\rho } ^2}}{{g_0^{\prime 2} + g_{\rho } ^2}}
\eeq
In the following analysis, we will take the same benchmark point
$g_\rho = 2.0$ as  in the last section, therefore $g_0 $ and
$g_0^\prime$  are fixed in order to  reproduce the SM model
couplings $g_1$ and $g_2$ at the electroweak scale. Including
higher order expansion of  the Higgs fields, the mixing would be
further modified, as indicated by the following derivation.
In the unitary gauge,  all the pion fields  are
eaten and  the Goldstone boson in the fourth direction is the
Higgs field, i.e.  $\pi^4 \simeq h^0 $. There is one interaction
term in the form of  $(h^0)^2 (g_0 \tilde{W}_\mu^a - g_0^\prime
\tilde{B}_\mu \delta^{a 3})$ embedded in the connection $E_\mu$'s
explicit expression :
\beq
E_{\mu}    &=& \left({g_0} \tilde{W}_\mu ^a T_L^a + g_0^\prime \tilde{B}_\mu T_R^3 \right) + \frac{1}{f^2} \left[ h^0 ~T^{\hat 4} ,[ { g_0 \tilde{W}_\mu^a  } T_L^a + g_0^\prime \tilde{B}_\mu T_R^3, h^0 ~T^{\hat 4} ] \right]+  \cdots  \nonumber \\
&=&  \left(g_0 \tilde{W}_\mu ^a T_L^a + g_0^\prime \tilde{B}_\mu T_R^3  \right)- \frac{(h^0)^2}{4 f^2} ( g_0 \tilde{W}_\mu ^a- g_0^\prime \tilde{B}_\mu \delta^{a 3} ) (T_L^a- T_R^a) + \cdots  \label{Emu}
\eeq
With the EW symmetry breaking, the second  term in the above equation gives rise to one new interaction term between the Higgs fields and gauge bosons.  

Substituting both $( h^0)^2 $ by their VEVs  would modify the mixing among the gauge bosons and vector mesons.  Assuming the charged $W_\mu^\pm$ gauge bosons are zero modes,  we will retain the correction occurring at the linear order of  $\xi$ but are justified to ignore corrections at the order of $m_W^2/m_\rho^2$. The full rotation for the charged gauge bosons is:
\beq
\tilde W_\mu ^ \pm  &=& \frac{g_\rho W_\mu^\pm + g_0 \rho _{L\mu }^\pm}{(g_\rho ^2 + g_0^2)^{1/2}} + \frac{ \xi g_\rho  g_0  (g_0 W_\mu^\pm - g_\rho \rho _{L\mu }^\pm)}{4 (g_\rho ^2 + g_0^2)^{3/2}} \\
\tilde \rho _{L\mu }^ \pm  &=& \frac{{g_0}{W_\mu^\pm } - {g_\rho }\rho _{L\mu }^\pm}{(g_\rho ^2 + g_0^2)^{1/2}} - \frac{ \xi g_\rho g_0 (g_\rho W_\mu^\pm + g_0 \rho _{L\mu }^\pm)}{4 (g_\rho ^2 + g_0^2)^{3/2}} - \frac{\xi}{4} \rho _{R\mu }^\pm \\
\tilde \rho _{R\mu }^ \pm  &=& \rho _{R\mu }^\pm + \frac{ \xi ( g_0 W_\mu^\pm  - g_\rho \rho _{L\mu }^\pm)}{ 4 (g_\rho ^2 + g_0^2)^{1/2}}
\eeq
where $\rho_L^\pm $ and $\rho_R^\pm $ are mass eigenstates with corresponding masses of $m_{\rho_L}^2 = g_\rho^2 a_\rho^2 f^2 $ and $ m_{\rho_R}^2 = (g_\rho^2+ g_0^2(1+ \xi/4) ) a_\rho^2 f^2 $.
The neutral gauge bosons mixing pattern is distinct from the charged ones as indicated by the $E_\mu$ expression [see Eq.(\ref{Emu})].  The eigenstates for the three massive neutral states are rather complicated.  However it is easy to project out the exact zero mode, i.e. the photon,  which is  the combination of the four neutral gauge eigenstates $\tilde{W}_\mu^3$, $\tilde{B}_\mu$, and $ \tilde{\rho}_{L \mu}^3$ , $\tilde{\rho}_{R \mu}^3$:
\beq
{A_\mu } &=& \frac{{g_0^\prime}{g_0}{{\tilde \rho }_{L\mu }} + {g_0^\prime}{g_0}{{\tilde \rho }_{R\mu }} + {g_\rho }({g_0^\prime}{{\tilde W}_\mu^3 } + {g_0}{{\tilde B}_\mu })}{\sqrt {g_\rho ^2(g_0^{\prime 2} + g_0^2) + 2g_0^{\prime 2} g_0^2} }  \eeq
and for completeness the Weinberg  mixing angle and the electromagnetic coupling are given as:
\beq
&& c_w^2  = \frac{g_0^2 (g_0^{\prime 2} + g_\rho^2)}{2 g_0^2 g_0^{\prime 2} + \left(
g_0^{\prime 2}+g_0^2\right) g_\rho^2} \approx \frac{g_0^2}{g_0^{\prime 2}+g_0^2} \\
&& s_w^2 = \frac{g_0^{\prime 2} (g_0^{2} + g_\rho^2)}{2 g_0^2 g_0^{\prime 2} + \left(
g_0^{\prime 2}+g_0^2\right) g_\rho^2} \approx  \frac{g_0^{\prime 2} }{g_0^{\prime 2}+g_0^2}  \\
&& e =\frac{g_0 g_0^\prime g_\rho}{\sqrt{2 g_0^2 g_0^{\prime 2}+ \left(g_0^{\prime 2}+g_0^2\right) g_\rho^2}} \approx \frac{ g_0 g_0^\prime }{ g_0^2 + g_0^{\prime 2}}
\eeq
which are consistent with the SM formulas as we abandon the corrections at the order of $g_0^\prime / g_\rho$ and $g_0 / g_\rho$. We prefer to conduct the calculation with the mass eigenstate  since the trilinear gauge  interaction  with one photon  and the quartic gauge  interaction  with  two photons are diagonal in that basis.

On the other hand,  with only one $h^0$ gain VEV in Eq. (\ref{Emu}) and  the mixing mass term gives us the following Lagrangian for $H$-$\tilde{\rho}$-$\tilde{W}$and $H$-$\tilde{\rho}$-$\tilde{B}$ interactions:
\beq
\mathcal{L}_{\rm{mix}} = \frac{ m_\rho^2 \xi}{2  g_\rho v}  h^0\tilde \rho _{L\mu }^a (g_0 \tilde W_\mu ^a - g_0^\prime \tilde B_\mu \delta^{a 3} )  -\frac{m_\rho^2 \xi }{2 g_\rho v}  h^0\tilde \rho _{R\mu }^a(g_0 \tilde W_\mu ^a - g_0^\prime \tilde B_\mu \delta^{a 3} ) \label{mix}
\eeq
Adapting it in terms of  the  mass eigenstates, we find positive shifts for the $h^0 W_\mu^+W_\mu^-$ and $h^0 Z_\mu^0 Z_\mu^0$ vertices and a negative shift for the  $h^0 \rho_{L \mu}^+ \rho_{L \mu}^-$ at the leading order of $\xi =v^2/f^2$. It is convenient to  parametrize  the Higgs interactions with the gauge bosons  adopting the effective theory approach:
\beq
\mathcal{L}_{eff} &=& a_W \frac{2m_W^2}{v} h^0 W_\mu ^+ W_\mu ^- + a_Z \frac{m_Z^2}{v} h^0 Z_\mu  Z_\mu  + c_\rho \frac{2 m_\rho^2}{v} h^0 \rho_{L \mu}^+ \rho_{L \mu}^- \nonumber \\  &+& c_{\rho_R W}  \frac{ m_\rho^2}{v} h^0 \left( W_\mu^+ \rho_{R \mu}^- +W_\mu^- \rho_{R \mu}^+\right)+ c_{\rho_L \rho_R}  \frac{ m_\rho^2}{v} h^0 \left(\rho_{L\mu}^+ \rho_{R\mu}^- + \rho_{L\mu}^- \rho_{R\mu}^+ \right) \nonumber \\  &+&  {c_f}  \left( \frac{m_f}{v}\bar f f \right) h_0 + c_\gamma \frac{\alpha }{8 \pi v } h^0 A^{\mu \nu } A_{\mu \nu } + c_{ Z\gamma} \frac{\alpha }{4 \pi v }  h^0 Z^{\mu \nu } A_{\mu \nu } \eeq
\beq
a_{W} &=& \left( \frac{g_\rho^2}{g_0^2 + g_\rho^2}+ \frac{g_0^2 g_\rho^2 \xi}{2 (g_0^2+g_\rho^2)^2} \right)\cos \theta+\frac{ g_0^2 \xi}{2 (g_0^2+ g_\rho^2)} \frac{m_\rho^2}{m_W^2} \\
c_{\rho } &=& \left( \frac{g_0^2}{g_0^2 + g_\rho^2} - \frac{g_0^2 g_\rho^2 \xi }{2 (g_0^2+g_\rho^2)^2} \right)\frac{m_W^2}{m_\rho^2} \cos \theta   - \frac{ g_0^2 \xi}{2(g_0^2+ g_\rho^2)}   \\
a_{Z} &=& \cos \theta+ \frac{\left(g_0^2 + g_0^{\prime 2} \right) m_\rho^2 \xi }{ 2 g_\rho^2 m_Z^2}
\eeq
where the third terms  in $a_W$ and $c_\rho$ and  the second term in $a_Z$ come from the mass mixing terms $h^0 \tilde{\rho}_L^a (g_0 \tilde{W}^a -g_0^\prime \tilde{B} \delta^{a 3})$ and  $h^0 \tilde{\rho}_R^a  (g_0 \tilde{W}^a -g_0^\prime \tilde{B} \delta^{a 3})$, and the $c_\gamma$ term is originated through the loop contribution of heavy charged particles. Notice that only diagonal  vertices are relevant to the  branching ratio of  Higgs decay into diphoton whereas there are additional nondiagonal Higgs vertices along with nondiagonal trilinear and quartic gauge interactions which would contribute to $h^0 \to Z \gamma$. The latter process is correlated to  $h^0 \to \gamma \gamma$ due to the electroweak symmetry. The corrections to $c_\gamma$ come from the vector meson and its mixing with $W$, $Z$ gauge bosons:
\beq
c_\gamma &=&  {c_t}{N_c} (2/3)^2F_{1/2}(4 m_t^2/ m_h^2) + a_{W} {F_1}(4 m_W^2/m_h^2) + c_{\rho}{F_1}(4 m_\rho^2/ m_h^2 )
\eeq
\beq
 && F_{1/2} (x) = -2 x \left(1+ (1 -x )\arcsin^2 (x^{-1/2} )\right) \\
 && F_1(x) =  2 + 3 x + 3 x  (2-  x){\arcsin^2 (x^{-1/2} )}
\eeq
with $ x_i =  4 m_i^2/ m_W^2$. For the large mass limit of $\rho$ mesons and top quark mass, the asymptotic values for those form functions are:  $F_{1}(x) \approx 7$ and $F_{1/2}(x) \approx -4/3$ .

\begin{figure}[h]
\begin{center}
\includegraphics[angle=0,clip,width=8cm]{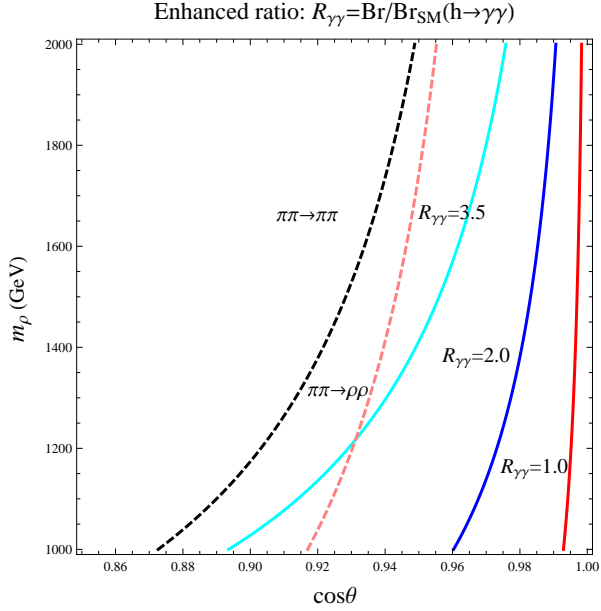}
\caption{{\small Contour plot for the $R_{\gamma \gamma}$ in the gluon fusion channel assuming $c_t =1$. The black dashed line is the unitary bound for the elastic pion scattering $\pi \pi \to \pi \pi$  and the orange dashed line is the unitary bound for the inelastic scattering $\pi \pi \to \rho \rho$ with $\Lambda = 5$~TeV. The region approaching the $\cos \theta = 1$ direction is permitted.}}
\label{photon_ratio}
\end{center}
\end{figure}
With the knowledge of those couplings, the partial width for Higgs to diphoton in the composite Higgs model with respect to its prediction in the SM and the respective ratios for  the other two bosons channels are fixed to be:
\beq
\Gamma/ \Gamma(H \to \gamma \gamma)^{sm} = \frac{c_\gamma^2 }{c_{\gamma ,sm}^2} , \Gamma/ \Gamma(H \to W W^*)^{sm} = \frac{a_W^2}{a_{W,sm}^2},  \Gamma/ \Gamma(H \to Z Z^*)^{sm} = \frac{a_Z^2}{a_{Z,sm}^2}
\eeq
However at  the LHC,  only the product of $\sigma \times Br (h \to VV^\prime) $ is measurable.The variable, which indicates the deviation of  composite Higgs models from the standard model, is the so-called $R$ parameter~\cite{Barducci}, i.e. the observing signal events divided by its corresponding SM expectation. For the diphoton process, the $R_{\gamma \gamma}$ is defined as:
\beq
 R_{\gamma \gamma} = \frac{\sigma \left( p p \to h^0 X \right) }{\sigma_{sm}\left( p p \to h^0 X \right) } \times \frac{Br (h^0 \to \gamma \gamma)}{ Br_{sm}( h^0 \to \gamma \gamma )}
\eeq
where $\sigma $ is the production cross section for  the Higgs boson
and $X $ denotes any particle associatively produced with the
Higgs boson. At the available energy scale, the main production channels for the Higgs bosons are gluon fusions $ g g \to h^0 $ and  vector boson fusions $ q \bar
q \to h^0 j j $. The modified cross sections for those two
processes are~\cite{Cacciapaglia}:
\beq
\frac{\sigma}{ \sigma_{sm}}( g g \to h^0 ) = c_t^2~, \quad
\frac{\sigma}{ \sigma_{sm}}( g g \to h^0 j j ) =  \frac{a_W^2 \sigma_{sm}^W + a_Z^2 \sigma_{sm}^Z}{\sigma_{sm}^W + \sigma_{sm}^Z}.
\eeq
For simplicity,  in this paper we are going to assume that all the fermion couplings are the same as they are in the SM  , i.e. $c_f = 1$, with the consequence that the top quark induced gluon fusion cross section is the same as in the SM. The observing ratios  for the diphoton process could be expressed in a more convenient form:
\beq
 R_{\gamma \gamma} = \frac{\sigma / \sigma_{sm} \cdot|c_\gamma /c_\gamma^{sm}|^2}{a_W^2 Br_{sm}^{(W W^*)} + a_Z^2 Br_{sm}^{(Z Z^*)}+|c_\gamma /c_\gamma^{sm}|^2 Br_{sm}^{(\gamma \gamma)} +\cdots}
\eeq
The $R_{\gamma \gamma}$ dependence on the $(\cos \theta, m_\rho)$
for the gluon fusion channel is plotted in
Fig.~\ref{photon_ratio}.  We put the unitary bound on that plot by
requiring that the perturbative unitary is violated at $\Lambda =5
$~TeV.  As we can see,  if we demand that the composite Higgs model
prediction does not give a significant deviation from the LHC
measurement,  the perturbative unitary is a very loose requirement
for the allowed parameter space.  To achieve a diphoton
enhancement rate not larger than a factor of 1.5,   we roughly need
$\cos \theta  >  0.97$ and $m_\rho > 1.0 $ TeV.  The $R_{\gamma
\gamma}$ in the vector boson fusion process is similar,  but with
$a_W, a_Z >1 $, a larger diphoton enhancement rate is encountered
in this channel.

Adding new fermion resonances to the composite model would be
quite interesting since, under certain circumstance, it possibly enhances the production cross section of Higgs bosons but at the same time it reduces the decay branching
ratio into diphotons. The balanced effect might depend on the
specific model details.  Furthermore, those composite fermions are
introduced into the model as vector-like quarks, thus their
mixing with the SM quarks would inevitably modify the $W$-$t$-$b$
and $Z$-$b$-$b$ vertices  and possibly give a  notable contribution
to the oblique parameters~\cite{Lavoura}. Detailed studies need to
be devoted to explore the influence of the third generation
composite quarks on the Higgs sector~\cite{Falkowski, Kearney, Carmona}.

\section{Conclusion}
In summary,  for a light composite Higgs boson which is
realized as one pNGB from a strong interacting sector,  $\pi \pi$ scatterings  put some mild
constraint on the $(\cos \theta, m_\rho)$ parameter space. We
conduct a careful analysis for both the elastic and inelastic pion
scatterings and  the deviation of  the Higgs to gauge couplings
from the standard model occurring at the  order of  $v^2/f^2$ is
allowed as we reduce  $m_\rho$, the mass of composite meson field.
The nonlinear realization enriches the Higgs interaction with SM
gauge bosons. It is noticed that in the minimal $SO(5)/SO(4)$
coset model, with the presence of  vector mesons in the
fundamental representation of $SO(4)$, a new interaction
originating from the strong interacting sector may shift the  Higgs
couplings $a_W$ and $a_Z$ in the positive direction  due to the
partial compositeness of $W$ and $Z$ gauge bosons after
electroweak symmetry breaking. Therefore it is easy for us to
accommodate an enhancement of diphoton rate which is observed at
the LHC.  It is believed that through extending the model structure (with effects on Higgs productions and decays) and fine tuning the parameter space, the light composite Higgs probably could fit the experimental
measurements much better than the standard model.

{\bf{Note}}: It is interesting to observe that there is nondiagonal contribution to  Higgs coupling to $Z$ and photon.  The calculation for this form factor is put in the appendix.

\section*{Acknowledgments}
The author is  grateful for discussion of the  manuscript with
Juan Sanz Cillero.  H.Cai is  supported by the postdoc foundation
under  Grant No. 2012M510001.

\newpage
\appendix
\section{non-diagonal gauge boson contribution to $H$-$Z$-$\gamma$}
In this appendix, we are going to show  that including  non-diagonal couplings exclusively results in a gauge-inviarant contribution for the form factor $c_{Z \gamma} $. Similar nondiagonal ccontribution from charginos in the MSSM is calculated in a reference \cite{Djouadi}.
\begin{figure}[h]
\begin{center}
\includegraphics[angle=0,clip,width=14 cm]{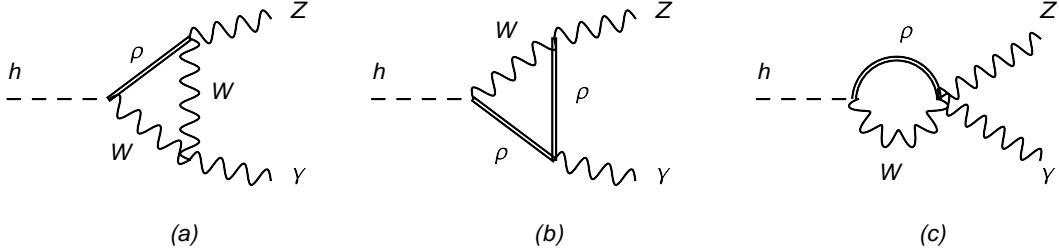}
\caption{{\small  (a) triangle feyndiagram with one vector resonance, (b) triangle feyndiagram with two vector resonances , (c) quartic feyndiagram with one vector resonance.}}
\label{feyn}
\end{center}
\end{figure} 

Assuming there is one non-diagonal Higgs-gauge coupling $c_{\rho_R W} \Frac{m_\rho^2}{v} \left(h^0 \rho_{R, \mu}^+  W_\mu^- + h.c. \right)$ and we can express some generic non-diagonal trilinear and quartic gauge self couplings in the following way: 
\beq
\mathcal{L}_{Z \rho_R W} &=&  (- i  \, e \, \cot \theta_w ) \, c_{Z \rho_R W} \bigg( \partial_\mu Z_\nu  \rho_{R, \mu}^+ W_\nu^- + \partial_\mu \rho_{R, \nu}^+ W_\mu^- Z_\nu + \partial_\mu W_\nu^- Z_\mu \rho_{R, \nu}^+   \nonumber \\ &+ &  \partial_\mu Z_\nu  \rho_{R, \nu}^- W_\mu^+ + \partial_\mu \rho_{R, \nu}^- W_\nu^+ Z_\mu + \partial_\mu W_\nu^+ Z_\nu \rho_{R, \mu}^-  - \left(\mu \leftrightarrow \nu \right)  \bigg) \\
\mathcal{L}_{A Z \rho_R W} &=&  (e^2 \, \cot \theta_w ) \, c_{Z \rho_R W} \bigg(2 ~ Z_\mu A_\mu  \rho^+_{R, \nu} W^-_\nu  -Z_\mu A_\nu \rho^+_{R, \mu} W^-_\nu  -Z_\nu A_\mu \rho^+_{R ,\mu} W^-_\nu  \nonumber  \\ 
&+ & 2 ~ Z_\mu A_\mu  \rho^-_{R,\nu} W^+_\nu  -Z_\mu A_\nu \rho^-_{R, \nu} W^+_\mu  -Z_\nu A_\mu \rho^-_{R, \nu} W^+_\mu \bigg)
\eeq
The amplitude for  Higgs decay into $Z$ and photon is adding up the three diagrams illustrated in Fig.[\ref{feyn}] and  it is necessary to times a factor of two to account for the crossing symmetry. As we put all the external particles to be on shell,  the amplitude can be organized into  a gauge invariant form:
\beq
&& M({h^0} \to Z\gamma ) = 2 \cdot \left (M^{(a)}+ M^{(b)} + M^{(c)}  \right) \nonumber  \\ 
&=&\frac{ - i \, e^2}{8 \pi ^2 v}  \left(2\, \, c_{\rho_R W} \, c_{Z\rho_R W} \right) \, c_{Z\gamma }^{(1)}  \, \left( {{g^{\mu \nu }}{k_1} \cdot {k_2} - k_1^\mu k_2^\nu } \right) \, \varepsilon _\mu ^*({k_1})\varepsilon _\nu ^*({k_2})
\eeq 
It is convenient to express the form factor $c_{Z \gamma}^{(1)} $ in terms of  one-loop three-point scalar and vector functions, i.e.  $C_0(m_h^2, m_Z^2, 0, m_2^2, m_1^2, m_2^2)$ and $C_2(m_h^2, m_Z^2, 0, m_2^2, m_1^2, m_2^2)$ defined in \cite{tHooft},  with two different masses circulating in the loop.
\beq
 c_{Z \gamma}^{(1)} &=& \cot \theta_w \cdot \frac{m_\rho^2}{ m_W^2} \cdot  \bigg[ \left(  \left( {\frac{m_h^2}{m_\rho^2} + \frac{m_W ^2}{m_\rho^2}  + 1 } \right) \left(m_Z^2 -m_\rho^2 -m_W^2 \right) -8 m_W^2 \right)   \nonumber \\ 
&\cdot &  \bigg( C_2 \left(m_h^2, m_Z^2,0,  m_W^2, m_\rho^2, m_W^2 \right) +C_2 \left(m_h^2, m_Z^2,0,  m_\rho^2, m_W^2, m_\rho^2 \right) \bigg)   \nonumber \\ 
 &+&   2 \left( \frac{m_W^2}{m_\rho^2} \left( m_Z^2 - m_W^2 - 3 m_\rho^2
\right) \right) \cdot C_0 \left(m_h^2, m_Z^2,0, m_W^2 ,m_\rho ^2 , m_W^2
\right)  \nonumber \\  
&+&   2 \left( m_Z^2 - 3 m_W^2 -m_\rho^2
\right) \cdot  C_0 \left(m_h^2, m_Z^2, 0, m_\rho ^2 ,m_W^2
,m_\rho ^2 \right) \bigg]
\eeq
where the special combination of  vector functions in the large bracket can be recasted into  Passarino-Veltman functions  $B_0$ and $C_0$:
\beq && C_2 \left(m_h^2, m_Z^2,0,  m_W^2, m_\rho^2, m_W^2 \right) +C_2 \left(m_h^2, m_Z^2,0,  m_\rho^2, m_W^2, m_\rho^2 \right) \nonumber \\ 
&=&   \frac{ m_Z^2 }{\left( {m_Z^2  - m_h^2 }
\right)^2}   \bigg( B_0 \left( {m_Z^2 ,m_\rho ^2 ,m_W^2 } \right) - B_0 \left( {m_h^2 ,m_\rho ^2 ,m_W^2 } \right)  \bigg)    \nonumber \\ 
&+&    \frac{1}{\left( {m_Z^2  - m_h^2 } \right)}  + \frac{m_W^2}{\left( m_Z^2-m_h^2 \right)} C_0\left( m_h^2 , m_Z^2,0 ,m_W^2, m_\rho^2 , m_W^2 \right)  \nonumber \\ & +& \frac{m_\rho^2}{\left(m_Z^2-m_h^2\right)}  C_0 \left(m_h^2, m_Z^2, 0, m_\rho ^2 ,m_W^2
,m_\rho ^2 \right) 
\eeq
It should be noticed that in the limit of $m_\rho = m_W$, our new form factor  $c_{Z \gamma}^{(1)}$ will reduce exactly to the $W$ gauge bosons mediated SM contribution\cite{hza}.

\end{document}